\documentclass[prd,draft,preprint,showpacs,groupedaddress]{revtex4-1}

\usepackage{amsmath}
\usepackage{amsfonts}
\usepackage{amssymb}
\usepackage{geometry}
\usepackage[english]{babel}

\begin{document}
\bibliographystyle{apsrev4-1}
\message{}

  \title{The generalization of charged AdS black hole specific~volume and number density }

  \author{Zi-Liang Wang} \author{Miao He} \author{Chao Fang}\author{Dao-Quan Sun}\author{Jian-Bo Deng} \email[Jian-Bo Deng: ]{dengjb@lzu.edu.cn}

  \affiliation{Institute of Theoretical Physics, LanZhou University,
    Lanzhou 730000, P. R. China}

  \date{\today}

  \begin{abstract}
  In this paper, by proposing a generalized $specific~volume$, we restudy the $P-V$ criticality of charged AdS black holes in the extended phase space. The results show that most of the previous conclusions can be generalized without change, but the ratio $\tilde{\rho}_c$ should be $3 \tilde{\alpha}/16$ in general case. Further research on the thermodynamical phase transition of black hole leads us to a natural interpretation of our assumption, and more black hole properties can be generalized. Finally, we study the number density for charged AdS black hole in higher dimensions, the results show the necessity of our assumption.
  
  \end{abstract}


  \maketitle

   \section{Introduction}
 
 Thermodynamics of black hole has been studied for many years, since the discovery of the four laws of black hole mechanics. And investigations have shown that black holes can not only have standard thermodynamic variables, but also possess rich phase structures and admit critical phenomena.    
 
  The phase transition of the Schwarzschild-AdS black hole was first demonstrated by Hawking and Page~\cite{a4}. Since then, the phase transitions and critical phenomena in different background have been studied \cite{a5,a6,a7,a8}. For charged AdS black hole,  it was found that the transition is of the Van der Waals (vdW) type in the canonical ensemble \cite{a7,a8}. Recently, by treating the cosmological constant as the pressure, $P=-\Lambda /8\pi$~\cite{a9,a10}, the complete analogy between the vdW fluid and charged AdS black hole was found, which is a great success in studying phase transition of black holes and  investigations in different black hole systems have been carried out\cite{a20,a21,a23,a24,a25,a26,a27,a28,a29,Liu:2016uyd,Majhi:2016txt,Zou:2013owa,Zou:2014mha,Zou:2016sab}. 
  
  Inspired by~\cite{a10}, the authors~\cite{aweiprl} introduced the number density of black hole molecules to measure the microscopic degrees of freedom of black holes. It can provide preliminary knowledge on the microscopic structure of a black hole and reveal novel interesting information.
  
  In this paper, we will generalize the $specific~volume$ identified in ~\cite{a10} , and restudy the $P-V$ criticality of charged AdS black holes in the extended phase space, showing that most of the previous conclusions can be generalized without change, but the ratio $\tilde{\rho}_c$ should be $3 \tilde{\alpha}/16$ in general case. Next, based on these results, we will investigate the black hole properties provided by the novel letter~\cite{aweiprl}.A natural interpretation of our assumption will be revealed, and we will have a deep understanding of the results in~\cite{aweiprl}. Finally, the generalization of number density for charged AdS black hole in higher dimensions will show the necessity of our assumption.

  This paper is organized as following. The basic equations of the black hole thermodynamics and $P-V$ criticality of charged AdS black holes in the extended phase space are summarized in Sec.~II. In Sec.~III, we generalize the $specific~volumn$ and restudy the results in \cite{a10}, then we extend our results to ~\cite{aweiprl} and study the number density in higher dimensions. Sec.~IV is reserved for conclusions and discussions.

\section{review of the thermodynamics and $P-V$ criticality of charged AdS black holes}

A black hole system was found to possess temperature $T$ and entropy $S$, i.e.~\cite{a1,a2,a3},
\begin{equation}
\label{1} T=\frac{\hbar \kappa}{2\pi c k_{B}}~, \ \ \ \ S=\frac{k_{B}c^3 A}{4 \hbar G}~,
\end{equation}
where $\kappa$ is the surface gravity and $A$ is the event horizon area of the black hole. In this paper, we use the units $\hbar = k_{B} = c = G=1$.  
And the first law of black hole thermodynamics is 
\begin{equation}
\label{2} dM=TdS+\Omega dJ +\Phi dQ~,
\end{equation}
where $\Omega$ is the angular velocity, $J$ is the angular momentum, $ \Phi$ is the electrostatic potential and Q is the electric charge. 

The phase transition of the Schwarzschild-AdS black hole was first demonstrated by Hawking and Page~\cite{a4}. Since then, the phase transitions and critical phenomena in different background have been studied \cite{a5,a6,a7,a8}. Recently, by treating the cosmological constant as the pressure, $P=-\Lambda /8\pi$~\cite{a9,a10}, the complete analogy between the vdW fluid and charged AdS black hole was found. Next, let us review the analogy.
 
 For charged AdS black hole, the first law of black hole thermodynamics in an extended phase space is \cite{Kastor:2009wy,a10}
 \begin{equation}
\label{3} dM=TdS +\Phi dQ + V dP~,
\end{equation}
where $V$ is thermodynamic volume, and in the simplest case of a Reissner-Nordstr$\ddot{\text o}$m-AdS (RN-AdS) black hole the volume is $V=4 \pi r_{+}^3/3$, with $r_{+}$ being the radius of the event horizon. From the metric of RN-AdS black hole in Schwarzschild-like coordinates, one can get the equation of state, 
 \begin{equation}
\label{4} P=\frac{T}{2r_{+}}- \frac{1}{8\pi r_+ ^{2}} + \frac{Q^2}{8\pi r_+ ^4}~.
\end{equation}
By dimensional analysis~\cite{a10}, the physical pressure and temperature are 
\begin{equation}
\label{5} {\text{Press}} = \frac{\hbar c}{l_P ^2}P~,\ \ \ \ {\text{Temp}}=\frac{\hbar c}{k}T,~
\end{equation}
and according to Eq.~\eqref{4}, one can get 
\begin{equation}
\label{6} {\text{Press}} = \frac{\hbar c}{l_P ^2}P= \frac{\hbar c}{l_P ^2} \frac{T}{2r_+}+\cdots= \frac{k{\text{Temp}}}{2l_P ^2 r_+} + \cdots ~.
\end{equation}
Comparing with the vdW equation, the authors~\cite{a10} identified the $specific~volume$
\begin{equation}
\label{7} v=2l_P^2 r_+ \  .
\end{equation}
With this identification, Eq.~\eqref{4} can be written as 
 \begin{equation}
\label{8} P=\frac{T}{v}- \frac{1}{2\pi v^{2}} + \frac{2Q^2}{\pi v^4}~.
\end{equation}
The  "$P-V$ diagram" is similar to the vdW gas~\cite{a10}. Moreover, the critical point can be obtained from 
 \begin{equation}
\label{9} \frac{\partial P}{\partial v}=0~,~\frac{\partial ^2 P}{\partial v^2}=0~,
\end{equation}
one can get
\begin{equation}
\label{10} T_c=\frac{\sqrt{6} }{18 \pi Q}~,~v_c=2 \sqrt{6} Q~,~ P_c=\frac{1}{96\pi Q^2}~,
\end{equation} 
Then, it leads to an interesting relation
\begin{equation}
\label{11} \rho_c \equiv \frac{P_c v_c}{T_c}=\frac{3}{8}~,
\end{equation} 
which is exactly the same as for the vdW fluid. For more novel results see ~\cite{a10}. 
This work is a great success in studying phase transition of black holes and investigations in different black hole systems have been carried out~\cite{a20,a21,a23,a24,a25,a26,a27,a28,a29}.

\section{The generalization of black hole specific volume and number density }
It is remarkable that the $\rho_c$ for RN-AdS black hole is the same as for the vdW fluid. It inspires us to wonder that whether this is the nature of black hole. In the first part of this section, we aim at finding out the answer. The result leads to the suggestion that it may be a special case of a more general situation. Next, we will show the derivation.

For later discussion, noticing the form of Eq.~\eqref{4}, we can suppose a general function
\begin{equation}
\label{12} P(x,T)=\frac{\alpha T}{x}+\frac{\beta }{x ^{2}} + \frac{\gamma}{x ^4}~,
\end{equation}
where $\alpha$, $\beta$ and $\gamma$ are constants.
 By the substitution 
\begin{equation}
\label{13} \tilde{x}=\tilde{\alpha} x ,
\end{equation} 
Eq.~\eqref{12} can be
\begin{equation}
\label{14} P(\tilde{x},T)=\frac{\alpha \tilde{\alpha}T}{\tilde{x}}+\frac{\tilde{\alpha}^2\beta }{\tilde{x} ^{2}} + \frac{\tilde{\alpha}^4\gamma}{\tilde{x }^4}~.
\end{equation}
One can get the "critical point" from
\begin{equation}
\label{15} \frac{\partial P}{\partial \tilde{x}}=0~,~\frac{\partial ^2 P}{\partial \tilde{x}^2}=0~,
\end{equation}
which leads to
\begin{equation}
\label{16} ~\tilde{x}_c=\tilde{\alpha}\sqrt{\frac{-6\gamma}{\beta}}~,~T_c=\frac{-4\beta}{3\alpha}\sqrt{\frac{-\beta }{6\gamma}}~,   ~ P_c=\frac{\beta ^2}{12\gamma }~,
\end{equation}
where one can find that $T_c$ and $P_c$ are independent of $\tilde{\alpha}$ (Physically, $ -\beta/\gamma$ and $-\beta/\alpha$ need to be nonnegative, luckily, this can be satisfied when we discuss the physical problem later.). 

Then, we have
\begin{equation}
\label{17} \tilde{\rho}_c \equiv \frac{P_c \tilde{x}_c}{T_c}=\frac{3}{8}\alpha \tilde{\alpha}~,
\end{equation} 
which means that $\tilde{\rho}_c$ only depends on $\alpha \tilde{\alpha}$.

So far, we have shown some characters of Eq.~\eqref{12} mathematically. One can find that Eq.~\eqref{12} will go back to Eq.~\eqref{4} when $\alpha=1/2$, $\beta=-1/8\pi$, $\gamma=Q^2/8\pi$ and $x=r_+$. So, from Eq.~\eqref{17}, it is obvious that we can get $\tilde{\rho}_c= 3/8$ with $\tilde{\alpha}=2$, which is the case (Eq.~\eqref{7}) we have shown in Sec.~II. In fact, the same form (Eq.~\eqref{12}) can also be found in different black hole systems~\cite{a20,a21}, and we can also obtain $\rho_c=3/8$ with $\tilde{\alpha}=1/\alpha$. 

Physically, the process of requiring $\tilde{\alpha}=1/\alpha$ in Eq.~\eqref{14} is inspired by the vdW equation

\begin{equation}
\label{vdW equation} (P+\frac{a}{v^2})(v-b)=T,
\end{equation}
i.e., identifying the $specific~volume$ $v=l_p^2 r_+ /\alpha$. Next, we will generalize this identification.

From Eq.~\eqref{6}, one can notice that $l_P ^2 r_+ $ has the dimension of volume, and without loss of generality, we argue that we should identify the $specific~volume$ $\tilde{v}$ with
\begin{equation}
\label{18} \tilde{v}= \tilde{\alpha} l_P ^2 r_+ .
\end{equation}

Returning to geometric units, the equation of state Eq.~\eqref{4} now is 
\begin{equation}
\label{19} P=\frac{\tilde{\alpha}T}{2\tilde{v}}- \frac{\tilde{\alpha}^2}{8\pi \tilde{v} ^{2}} + \frac{\tilde{\alpha}^4Q^2}{8\pi \tilde{v} ^4}~.
\end{equation}
From Eq.~\eqref{16},the critical point in this case is 
\begin{equation}
\label{20} ~\tilde{v}_c=\tilde{\alpha}\sqrt{6}Q~, T_c=\frac{\sqrt{6} }{18 \pi Q}~,~ P_c=\frac{1}{96\pi Q^2}~.
\end{equation} 
 As we have mentioned before, $T_c$ and $P_c$ are independent of $\tilde{\alpha}$, so they are the same with Eq.~\eqref{10}. 
 
 And one can have
 \begin{equation}
\label{21} \tilde{\rho}_c =\frac{3}{16}\tilde{\alpha}~.
\end{equation} 
which only depends on $\tilde{\alpha}$, and for a specific $\tilde\alpha$, $\tilde{\rho}_c$ is a universal number for any RN-AdS black hole with arbitrary charge.

Following~\cite{a10}, one can define 
\begin{equation}
\label{22} p=\frac{P}{P_c}~,~\nu=\frac{\tilde{v}}{\tilde{v}_c}~,~\tau =\frac{T}{T_c},
\end{equation} 
then, Eq.~\eqref{19} can be written as 
\begin{equation}
\label{23} 8\tau = 3\nu ~(p+\frac{2}{\nu ^2})-\frac{1}{\nu ^3},
\end{equation} 
which is independent of $\tilde{\alpha}$, so you can find the same equation in~\cite{a10}, and it can cause the same critical exponents. 

Since the Gibbs free energy is a function of $P$ and $T$, it is still the same in~\cite{a10}. 

So far, by supposing a general $specific~volume$ of RN-AdS black hole, we restudied the results in~\cite{a10}. The results show that most of the previous conclusions can be generalized without change, but the ratio $\tilde{\rho}_c$ should be $3 \tilde{\alpha}/16$ in general case.
So, our generalization would be practicable. Next, based on these results, we will investigate more about thermodynamical phase transition.

In~\cite{aweiprl}, to measure the microscopic degrees of freedom of black holes, the number density of black hole molecules was introduced, which is defined as
\begin{equation}
\label{24} n=\frac{1}{v}=\frac{1}{2l_P ^2 r_+ },
\end{equation} 
and the authors had given a natural interpretation for it from the holographic view~\cite{aweiprl,a31,a41}. The microscopic degrees of freedom of the black hole was proposed to be carried by the Planck area pixels. By assuming that one microscopic degree of freedom occupies $\lambda$ Planck area pixels, the total number of the microscopic degrees of freedom is 
\begin{equation}
\label{25} N=\frac{A}{\lambda l_P ^2},
\end{equation}
and for RN-AdS black hole, one can have
\begin{equation}
\label{26} n=\frac{N}{V}=\frac{3}{\lambda l_P ^2 r_+},
\end{equation}
where $V$ is the thermodynamic volume. To obtain Eq.~\eqref{25}, the authors took $\lambda=6$, which means that one microscopic degree of freedom occupies 6 Planck area pixels. This result is interesting, since we did not confirm the definite number before, for example, Bekenstein~\cite{a41} proposed that the number would be 4. One may wonder that whether this is actually the microscopic structure of AdS black hole. In fact, our results may give a hint to the answer.

Similar to Eq.~\eqref{24}, based on the general $specific~volume$ we proposed, one can define the general number density
\begin{equation}
\label{27} \tilde{n} =\frac{1}{\tilde{v}}=\frac{1}{\tilde{\alpha}l_P ^2 r_+ }.
\end{equation}
It is easy to find that this equation is just Eq.~\eqref{26} with the replacement $\tilde{\alpha}=\lambda /3$. This gives the $\tilde{\alpha}$ in our assumption Eq.~\eqref{18} a natural physical meaning, it is corresponding to the the number of Planck area pixels that one microscopic degree of freedom occupies. So, once you chose a specific $\tilde{\alpha}$, you will get a specific $ \lambda$, for example, $\lambda=6$ when $\tilde{\alpha}=2$. 

Moreover,  all the results in ~\cite{aweiprl} are based on the relation between $T/T_c$ and $n/n_c$ or the relation between $P/P_c$ and $n/n_c$~. From Eq.~\eqref{20}, we know that $T_c$ and $P_c$ are independent of $\tilde{\alpha}$, and 
\begin{equation}
\label{28} \frac{n}{n_c}=\frac{v_c}{v}=\frac{2l_P ^2 r_c}{2l_P ^2 r_+}=\frac{\tilde{\alpha}l_P ^2 r_c}{\tilde{\alpha}l_P ^2 r_+ } =\frac{\tilde{v}_c}{\tilde{v}}=\frac{\tilde{n}}{\tilde{n}_c},
\end{equation}
which is also independent of $\tilde{\alpha}$. So, all the interesting results hold for any $\tilde{\alpha}$, in other words, one can get the same conclusions in~\cite{aweiprl} with the general number density we defined (For example, the general number density suffers a sudden change accompanied by a latent heat when
the black hole system crosses the small-large black
hole coexistence curve.). 

As one can see, once our assumption Eq.~\eqref{27} is accepted, there is no constrain on the number of Planck area pixels that one microscopic degree of freedom occupies from the thermodynamical phase transition of black hole. Maybe we should determine the definite $ specific~volume $ of charged AdS black hole from the microscopic structure of black hole.

If one apply the holographic view to the $specific~volume$ of charged AdS black hole in higher dimensions ($d > 4$), the necessity of our assumption will be revealed.
 
 In this case, the equation of state is~\cite{a50} 
\begin{equation}
\label{30} P=\frac{T(d-2)}{4 r_+}-\frac{(d-3)(d-2)}{16\pi r_+ ^2}+\frac{q^2(d-3)(d-2)}{16\pi r_+ ^{2(d-2)}},
\end{equation}
where $q$ is related to the black hole charge $Q$. The original $specific ~volume$ is identified as
 \begin{equation}
\label{31} v=\frac{4r_+ l_P ^{d-2}}{d-2}.
\end{equation}
So, the corresponding number density is 
 \begin{equation}
\label{32} n=\frac{d-2}{4r_+ l_P ^{d-2}}.
\end{equation} 
And the thermodynamic volume and the event horizon area are~\cite{a50} 
 \begin{equation}
\label{33} V=\frac{\omega _{d-2}r_+ ^{d-1}}{d-1}~,~ A_{d-2}= \omega _{d-2}r_+ ^{d-2},
\end{equation}
where $ \omega _{d}$ is the volume of the unit
$d-$ sphere,
\begin{equation}
\label{34} \omega_{d}= \frac{2\pi ^{\frac{d+1}{2}}}{\Gamma (\frac{d+1}{2})}.
\end{equation}
From holographic view, the total number of the microscopic degrees of freedom is

\begin{equation}
\label{35} N=\frac{A_{d-2}}{\lambda l_P ^{d-2}},
\end{equation}
then,
\begin{equation}
\label{36} n=\frac{N}{V}=\frac{d-1}{\lambda l_P ^{d-2} r_+}.
\end{equation}
 
Combining Eq.~\eqref{32} and Eq.~\eqref{36}, one can get
\begin{equation}
\label{37} \lambda=\frac{4(d-1)}{d-2}.
\end{equation}
One can find that the $\lambda$ will not be an integer except for $d=6$, for example, in $5$ dimension, $\lambda$ is $16/3$, which is improper since the basic unit is $l_P ^{d-2}$ and $\lambda$ should be integer. 

This problem can be avoidable with our generalized  $specific~volume$. In this case, since $r_+ l_P ^{d-2}$ has the dimension of volume in $d$ dimension, the general $specific~volume$ should be 
\begin{equation}
\label{38} \tilde{v}=\tilde{\alpha}r_+ l_P ^{d-2},
\end{equation}
 and the corresponding number density is 
 \begin{equation}
\label{39} \tilde{n}=\frac{1}{\tilde{\alpha}r_+ l_P ^{d-2}}.
\end{equation} 
Combining Eq.~\eqref{39} and Eq.~\eqref{36}, we have
 \begin{equation}
\label{40} \tilde{\alpha}=\frac{\lambda}{d-1}.
\end{equation}
So, $\lambda$ can also be integer in higher dimensions. In other words, the number density corresponding to the general $specific~volume$ is well defined in higher dimensions.  


\section{conclusions and discussions}
To summarize, in this paper, by supposing a general function Eq.~\eqref{12}, we first found that the $\tilde{\rho}_c=3 \alpha \tilde{\alpha}/8$,
which only depends on $\alpha \tilde{\alpha}$, and one can always obtain $\tilde{\rho}_c=3/8$ by choosing $ \tilde{\alpha}= 1/\alpha$. Motivated by this, we generalized the $specific ~ volume$~$\tilde{v}=\tilde{\alpha} l_P ^2 r_+ $  based on the dimensional analysis, and then, we restudied the results in original paper ~\cite{a10}, we found that the critical $specific ~ volume$ $~\tilde{v}_c=\tilde{\alpha}\sqrt{6}Q~$ and $\tilde{\rho}_c=3 \tilde{\alpha}/16$, which can go back to the case in~\cite{a10} with $\tilde{\alpha} =2$. Calculations revealed that $P_c$ , $T_c$ and "the law of corresponding states" (Eq.~\eqref{23}) are independent of $\tilde{\alpha}$. We also found that the Gibbs free energy and the critical exponents can be generalized without change. All these results show the practicability of our assumption.

Based on these results, we investigated the black hole properties provided by the Letter~\cite{aweiprl}. In this case, by defining a  general number density, we found that the $\tilde{\alpha}$ in our assumption of $specific ~ volume$ is corresponding to the the number of Planck area pixels that one microscopic degree of freedom occupies, so it adds a natural physical interpretation of our assumption. Moreover, calculations revealed that all the novel results in ~\cite{aweiprl} still  hold for the general number density we defined, for example, the general number density suffers a sudden change accompanied by a latent heat when the black hole system crosses the small-large black hole coexistence curve.

Further,  if our assumption is accepted, there is no constrain on the number of Planck area pixels that one microscopic degree of freedom occupies from the thermodynamical phase transition of black hole, which is different from~\cite{aweiprl}. This would be reasonable, since we still can not confirm the definite number. 

Finally, we studied the situation for charged AdS black hole in higher dimensions. We found that the number density~\cite{aweiprl} can also have a natural interpretation from holographic view, but the corresponding $specific ~ volume$ must be our general one.

We close by noting that we have only restudied the $P-V$ criticality of charged black hole with the generalized $specific ~volume$. Since our generalization is based on dimensional analysis, it can extend to different black hole systems and more general properties will be revealed. It is unknown whether this kind of generalizations in other different black hole systems can also have a natural interpretation. All these are left for the further research.

\section*{ACKNOWLEDGMENTS}

  We would like to thank the National Natural Science Foundation of
  China~(Grant No.11171329) for supporting us on this work.

  \bibliographystyle{apsrev4-1}
  \bibliography{reference}

\end{document}